# Linear Analog Codes: The Good and The Bad

Kai Xie,   and   Jing Li (Tiffany)

Electrical and Computer Engineering Department, Lehigh University, Bethlehem, PA 18015

Email: {kax205, jingli}@ece.lehighe.edu

*Abstract*—This paper studies the theory of linear analog error correction coding. Since classical concepts of minimum Hamming distance and minimum Euclidean distance fail in the analog context, a new metric, termed the "minimum (squared Euclidean) distance ratio," is defined. It is shown that linear analog codes that achieve the largest possible value of minimum distance ratio also achieve the smallest possible mean square error (MSE). Based on this achievability, a concept of "maximum distance ratio expansible (MDRE)" is established, in a spirit similar to maximum distance separable (MDS). Existing codes are evaluated, and it is shown that MDRE and MDS can be simultaneously achieved through careful design.

## I. INTRODUCTION

In a typical digital communication system, analog source samples (such as sound and images) are first quantized and labeled to binary sequences, then encoded by a digital error correction code (DECC), and finally modulated by a digital modulator before being sent over the channel. However, quantization inevitably introduces irrecoverable granularity error, as well as significantly increases the data volume. Additionally, quantization also causes the issue of "most/more significant bits" vs "least/less significant bits", and would in general require judicious unequal error protection (UEP) to avoid wasteful protection of some bits while inadequate protection of others. An alternative to this cumbersome, although rather commonplace, digital approach is to completely rid of quantization, directly encode analog data sequences to analog codewords, and transmit these analog symbols over the channel as in an $\infty$-order amplitude shift keying (if real-valued) or quadratic amplitude modulation (if complex-valued). As such, a single *analog error correction code (AECC)* can replace the combination of quantization, DECC and digital modulation, and be free of quantization error floor.

The notion of analog error correction is not new. It traces back to the early 80's, when Marshall and Wolf independently introduced the concept [1]–[3]. It was termed *real number coding* in Marshall's work and *analog coding* in Wolf's work. Early work of AECC presents a natural outgrowth of linear digital codes, by extending conventional linear digital codes from the finite field to the real- or complex-valued field (symbols from a very large finite field can approximate real values). Hence, linear codes prevail the short literature of analog coding, just as they do in digital coding. There have also been proposals for nonlinear analog codes and some of them actually exhibited surprisingly good performances [8], [9]. In general, however, the research of analog coding is still very limited, especially comparing to the high level of maturity of digital coding in both theoretical and practical contexts.

It is apparent that analog codes generalize digital codes by relaxing the source space and the codeword space from discrete fields containing finite elements to continuous fields containing (uncountably) infinite elements [4]. However, it is less apparent what fundamental or subtle implications analog codes cast – especially in terms of coding theory and practical code design, compared to the usual practice of digital coding. Intuitively, Hamming distance, a concept of critical importance to digital codes, is much less indicative in analog codes, since two analog sequences can have a large Hamming distance but still be similar to each other (e.g. every symbol differs, but very minorly), or have a small Hamming distance but be far apart (e.g. only one symbol differs, but the difference is huge). However, Euclidean distance does not appear to be a good metric either. As we will show, the minimum Euclidean distance of any analog code is infinitesimal, and hence analog codes do not have a guaranteed error correction capability as digital codes do. In the presence of additive white Gaussian noise (AWGN), every decoded frame of an analog code is bound to contain non-zero error/distortion, leading to an word error rate of always 1. All of this departs from the conventional digital coding theory, and awaits to be illuminated.

This paper presents a theoretic study of analog codes, and *linear* codes in particular. We develop several new concepts for analyzing and understanding linear analog codes, including the *encoding power gain*, the *minimum (squared Euclidean) distance ratio* and its achievable upper bound, and the mean square error (MSE) distortion and its achievable lower bound. We establish a concept of *maximum distance ratio expansible (MDRE)* for linear analog codes, which is close in spirit to maximum distance separable (MDS) in linear digital codes. We show that MDRE codes can achieve the best (i.e. largest) minimum distance ratio and the best (i.e. smallest) MSE distortion. In this, we show that all MDRE codes provide the same, best MSE distortion on AWGN channels. We identify analog codes that are MDRE (as well as MDS), and show that the criteria of MDRE and MDS, although evaluated against different distance metrics, need not conflict each other, but can be effectively unified in the same code design. We also evaluate existing linear analog codes and MDRE codes in particular. One important notion developed here is, unlike in digital coding where linear codes are sufficient to combat Gaussian noise, linear analog codes are actually rather weak and inadequate, and hence *nonlinear* transforms seem necessary, in order to provide performance that will be appreciated in practical applications.

[0]Supported by National Science Foundation under Grants No. 0928092, 0829888 and 1122027.

## II. PRELIMINARY

In this paper, we use bold fonts, such as $\mathbf{G}$ and $\mathbf{u}$, to denote matrices or vectors (column vectors by default), and use regular fonts, such as $n$ and $\Gamma$ to denote scalars. Superscript $^T$ denotes simple transpose of a vector or matrix, while superscript $^H$ denotes the Hermitian transpose. By default, all the analog codes have parameters $(n,k)$, and maps a length-$k$ discrete-time complex-valued source sequence $\mathbf{u} = (u_0, u_1, \cdots, u_{k-1})^T \in \mathbb{C}^k$ to a length-$n$ discrete-time complex-valued codeword $\mathbf{v} = (v_0, v_1, \cdots, v_{n-1})^T \in \mathbb{C}^n$. Since any linear code can be expressed in the form of a linear block code, we focus the discussion on linear block codes. Similar to digital codes, an analog linear block code is completely specified by its *generator matrix*, $\mathbf{G}_{k \times n}$, a rank-$k$ real-valued matrix satisfying $\mathbf{v}(\mathbf{u}) = \mathbf{G}^H \mathbf{u}$. After the codeword $\mathbf{v}$ passes though a channel with additive noise $\mathbf{w}$, the decoder receives $\mathbf{r} = \mathbf{v} + \mathbf{w}$, and produces an estimate $\tilde{\mathbf{u}}$ of the original source vector $\mathbf{u}$.

Before we proceed, we first quickly overview the existing linear analog codes. The first, and one of the most important class, is *discrete Fourier transform* (DFT) codes, due to Marshall [1] and Wolf [2]. The generator matrix of an $(n,k)$ DFT code is formed by extracting a set of $k$ rows from the (normalized) DFT matrix $\boldsymbol{\Psi}$ of order $n$, where each element $\Psi_{i,k} \in \boldsymbol{\Psi}$ id defined as $\forall i = 0, 1, \cdots, n-1, k = 0, 1, \cdots, n-1$,

$$\Psi_{i,k} = \frac{1}{\sqrt{n}} \left( e^{-j2\pi/n} \right)^{2ik}. \quad (1)$$

When the extracted rows follow certain structural formalism, the resultant complex DFT code can be viewed as an analog Bose-Chaudhuri-Hocquenghem (BCH) code and at the same time satisfies maximum distance separability [3]. In other words, there exist a subclass of DFT codes that are by nature analog Read-Solomon (RS) codes and hence optimal in the MDS sense. It has been shown that the traditional decoders of digital BCH codes, such as Peterson-Gorenstein-Zierler (PGZ) decoder, Berlekamp-Massey algorithm and Forney algorithm, are also applicable to analog BCH codes (and are useful when the channel is an erasure channel or a pulse channel).

Another important class of analog codes, which happen to be also MDS, are *discrete cosine transform* (DCT) codes due to Wu and Shiu [5]. Similar to DFT codes, the generator matrix $\mathbf{G}$ of a DCT code comprises $k$ selected rows from a DCT matrix $\boldsymbol{\Xi}$, where each element $\xi_{i,k} \in \boldsymbol{\Xi}$ is defined as

$$\xi_{i,k} = \begin{cases} 1/\sqrt{n} & k=0 \\ \frac{2}{\sqrt{n}} \cos \frac{(2i+1)k\pi}{2n} & k=1,2,...,n-1 \end{cases}. \quad (2)$$

Different from DFT codes, DCT codes are not analog BCH codes or even cyclic codes. However, since the parity check matrix preserves the properties of a Vandermonde matrix, DCT codes are nevertheless MDS. Additionally, a specific subclass of DCT codes can be expressed in a BCH-like structure and decoded by a modified Berlekamp-Massey and Forney algorithm [5]. This BCH-like DCT structure was later generalized to *discrete sine transform* (DST) codes by Rath and Guillemot [6], and a subspace-based decoder is proposed for general DCT and DST codes in [7].

We note that (normalized) DFT codes and DCT/DST codes are all *unitary codes*, i.e. codes whose generator matrix is formed by a selected set of rows from a square unitary matrix.

## III. STRUCTURE PROPERTIES: DISTANCE RATIO & MDRE

Clearly, generator matrices $\mathbf{G}$ and $a\mathbf{G}$ ($a > 1$) define essentially the same code, and the seemingly larger distance expansion of the latter is only the artifact of a larger consumption of transmission energy. To facilitate a fair comparison, we introduce the definition of encoding power gain.

*Definition 1:* The *encoding power gain* $\Gamma$ of a generator matrix $\mathbf{G}$ is defined as the ratio between the average codeword power and the average source vector power:

$$\Gamma \triangleq \frac{\int P(\mathbf{u}) \mathbf{v}^H \mathbf{v} d\mathbf{u}}{\int P(\mathbf{u}) \mathbf{u}^H \mathbf{u} d\mathbf{u}} \quad (3)$$

where $P(\mathbf{u})$ is the probability density function (pdf) of the source vector $\mathbf{u}$, and $\int f(\mathbf{v}) d\mathbf{v}$ represents the multiple integrals $\int \int \cdots \int f(v_0, v_1, \cdots, v_{n-1}) dv_{n-1} \cdots dv_1 dv_0$. An analog linear block code is said *normalized*, if $\Gamma = 1$.

*Theorem 1:* Consider an analog linear block code with generator matrix $\mathbf{G}$. The source vector $\mathbf{u}$ consists of $k$ elements $u_i$, each drawn from the same i.i.d (independent and identically distributed) source with distribution $p$. The encoding power gain is given by $\Gamma = trace(\mathbf{GG}^H)/k$.

Proof:
$$\Gamma = \frac{\int P(\mathbf{u}) \mathbf{v}^H \mathbf{v} d\mathbf{u}}{\int P(\mathbf{u}) \mathbf{u}^H \mathbf{u} d\mathbf{u}} = \frac{\int P(\mathbf{u}) \mathbf{u}^H \mathbf{GG}^H \mathbf{u} d\mathbf{u}}{\int P(\mathbf{u}) \mathbf{u}^H \mathbf{u} d\mathbf{u}} \quad (4)$$

Since $\mathbf{GG}^H$ is Hermitian and positive definite, it is possible to perform a singular value decomposition, such that $\mathbf{GG}^H = \mathbf{A}^H \mathbf{DA}$, where $\mathbf{A}$ is a square unitary matrix and $\mathbf{D}$ is a real-valued diagonal matrix with positive diagonal elements $\{d_0, d_1, ..., d_{k-1}\}$. We have

$$\begin{aligned} \Gamma &= \frac{\int P(\mathbf{u}) \mathbf{u}^H \mathbf{A}^H \mathbf{DAu} d\mathbf{u}}{\sum_{i=0}^{k-1} \int p(u_i) |u_i|^2 du_i} = \frac{\sum_{i=0}^{k-1} (d_i \int p(u_i) |u_i|^2 du_i)}{k \int p(u_0) |u_0|^2 du_0} \\ &= \frac{\int p(u_0) |u_0|^2 du_0 \left( \sum_{i=0}^{k-1} d_i \right)}{k \int p(u_0) |u_0|^2 du_0} = \frac{\sum_{i=0}^{k-1} d_i}{k} \\ &= \frac{trace(\mathbf{GG}^H)}{k}. \end{aligned} \quad (5)$$

*Corollary 2:* Unitary codes have encoding gain $\Gamma = k/k = 1$, and are therefore normalized.

An error correction code provides error protection by expanding the distances among sequences. On AWGN channels, squared Euclidean distance becomes very relevant, since it constitutes the exponential part of the Gaussian distribution, and is closely related to the likelihood test. The squared Euclidean distance of two codewords $\mathbf{v}$ and $\mathbf{v}'$ is given by $d_E^2(\mathbf{v}, \mathbf{v}') = ||\mathbf{v} - \mathbf{v}'||^2 = \sum_{i=0}^{n-1} |v_i - v_i'|^2$.

Due to the geometric uniformity of linear codes, the all-zero sequence is not only always a valid codeword, but can also act as a typical codeword in term of distance analysis. Just like Hamming distance spectrum and Hamming weight spectrum are used interchangeably in linear digital codes, squared Euclidean distance spectrum and squared Euclidean weight spectrum are the same one in linear analog codes. The squared Euclidean weight of a codeword $\mathbf{v}$ is given by $w_E^2(\mathbf{v}) = d_E^2(\mathbf{v}, \mathbf{0})$.

*Theorem 3:* The minimum squared Euclidean distance of an analog linear block code always approaches 0.

Proof: It is sufficient to show that, for an arbitrarily small positive value $\varepsilon$, there exists a codeword $\mathbf{v} = \mathbf{Gu}$ whose squared Euclidean weight $w_E^2(\mathbf{v}) < \varepsilon$. Consider a source sequence $\mathbf{u} = (u_0, 0, ..., 0)^T$ having only one non-zero element $u_0$. The corresponding codeword $\mathbf{v}$ has weight:

$$w_E^2(\mathbf{v}) = \sum_{i=0}^{n-1} |v_i|^2 = ||\mathbf{G}^H \mathbf{u}||^2 = |u_0|^2 \sum_{i=0}^{n-1} |g_{i0}|^2, \quad (6)$$

where $g_{ij}$ is the element in the $i$th row and $j$th column of $\mathbf{G}$. Thus, if we select $u_0$ to be a real positive number satisfying

$$0 < u_0 < \sqrt{\frac{\varepsilon}{\sum_{j=0}^{n-1} g_{0j} * g_{0j}}}, \quad (7)$$

the corresponding codeword $\mathbf{v}(\mathbf{u})$ has a squared Euclidean weight $w_E^2$ smaller than $\varepsilon$.

Since the minimum Euclidean distance/weight of analog linear codes can be arbitrarily small, it can no longer indicate the structural goodness of the code. Instead, we introduce a new metric, the *distance ratio*.

*Definition 2:* Consider a pair of source sequences $\mathbf{u}$ and $\mathbf{u}'$ and their respective codewords $\mathbf{v}$ and $\mathbf{v}'$. The *squared Euclidean distance ratio*, or simply, the *distance ratio* between them is defined as

$$\mathcal{R}_E^2(\mathbf{u}, \mathbf{u}') = \frac{d_E^2(\mathbf{v}, \mathbf{v}')}{d_E^2(\mathbf{u}, \mathbf{u}')} = \frac{||\mathbf{v} - \mathbf{v}'||^2}{||\mathbf{u} - \mathbf{u}'||^2}. \quad (8)$$

The smallest distance ratio among all the source pairs is termed the *minimum (squared Euclidean) distance ratio* of the code. The *average (squared Euclidean) distance ratio* of the code is defined as

$$\bar{\mathcal{R}}_E^2 = \int P(\mathbf{u}) \mathcal{R}_E^2(\mathbf{u}, \mathbf{0}) d\mathbf{u}. \quad (9)$$

*Theorem 4:* For all linear analog codes with encoding power gain $\Gamma$, their minimum distance ratio is upper bounded by $\Gamma$, and the upper bound is achieved by the generator matrix $\mathbf{G}$ such that all the $k$ eigenvalues of $\mathbf{GG}^H$ are identical.

Proof: It is sufficient to consider the distance ratio between an arbitrary non-zero codeword $\mathbf{v}(\mathbf{u})$ and the all-zero codeword.

$$\mathcal{R}_E^2(\mathbf{u}, \mathbf{0}) = \frac{||\mathbf{v}||^2}{||\mathbf{u}||^2} = \frac{\mathbf{u}^H \mathbf{GG}^H \mathbf{u}}{\mathbf{u}^H \mathbf{u}}. \quad (10)$$

Decompose $\mathbf{GG}^H$ to $\mathbf{A}^H \mathbf{DA}$, where $\mathbf{A}$ is a unitary matrix and $\mathbf{D}$ is a diagonal matrix with positive diagonal elements $\{d_0, d_1, ..., d_{k-1}\}$. Let $d_{min}$ be the smallest of all the diagonal elements: $d_{min} = \min\{d_0, d_1, ..., d_{k-1}\} > 0$. We can simplify (10) to

$$\mathcal{R}_E^2(\mathbf{u}, \mathbf{0}) = \frac{\mathbf{u}^H \mathbf{A}^H \mathbf{DAu}}{\mathbf{u}^H \mathbf{u}} \geq d_{min} \frac{\mathbf{u}^H \mathbf{A}^H \mathbf{IAu}}{\mathbf{u}^H \mathbf{u}} \quad (11)$$

$$= d_{min} \frac{\mathbf{u}^H \mathbf{A}^H \mathbf{Au}}{\mathbf{u}^H \mathbf{u}} = d_{min} \frac{\mathbf{u}^H \mathbf{u}}{\mathbf{u}^H \mathbf{u}} = d_{min}, \quad (12)$$

where $\mathbf{I}$ is an identical matrix. The source vector $\mathbf{u}$ that achieves the equality in (11) is one that satisfies $\mathbf{u}' = \mathbf{Au} = (0, 0, \cdots, u_i, \cdots, 0)^T$, where $i$ is the index for $d_{min}$.

Further,

$$\min\left(\mathcal{R}_E^2(\mathbf{u}, 0)\right) = d_{min} \leq \frac{\sum_{i=0}^{k-1} d_i}{k} \quad (13)$$

$$= \frac{trace(\mathbf{GG}^H)}{k} = \Gamma. \quad (14)$$

The equality in (13), i.e. the upper bound of the minimum distance ratio, is achieved when all the eigenvalues of $\mathbf{GG}^H$ are identical: $d_1 = d_2 = \cdots = d_{k-1} = d_{min}$.

*Corollary 5:* Given an $(n, k)$ linear analog code with generator matrix $\mathbf{G}$, its minimum distance ratio $d_{min}$ is the smallest eigenvalue of the matrix $\mathbf{GG}^H$.

*Definition 3:* Consider all the linear analog codes with a fixed encoding power gain $\Gamma$. A code is called *maximum distance ratio expansible* or *MDRE*, if its minimum distance ratio achieves the upper bound $\Gamma$ with equality.

*Corollary 6:* An $(n, k)$ analog linear block code with generator matrix $\mathbf{G}$ is MDRE, if and only if all the $k$ eigenvalues of $\mathbf{GG}^H$ are identical.

*Theorem 7:* Analog unitary codes are MDRE.

Proof: For an analog unitary code, we have $\mathbf{GG}^H = \mathbf{I}$. That is, $\mathbf{GG}^H$ has identical eigenvalues (i.e. 1) and the code is therefore MDRE.

Since (normalized) DFT codes and DCT/DST codes are all unitary codes, they are also MDRE. What is less expected is that repetition codes are also MDRE.

*Theorem 8:* Analog repetition codes are MDRE.

Proof: Consider an analog repetition code that repeats the length-$k$ source vector $t$ times. The generator matrix consists of $r$ identity matrices of rank $t$ each: $\mathbf{G} = [\mathbf{I}_k, \mathbf{I}_k, \cdots, \mathbf{I}_k]$. Since $\mathbf{GG}^H = t\,\mathbf{I}_k$, all the eigenvalues are identical (i.e. $t$), and the code is therefore MDRE.

Since MDRE codes do best in terms of distance expansion (given the same encoding power gain), it is reasonable to expect them to perform well – at least better than the rest of linear analog codes. However, the fact that analog repetition codes are MDRE, yet their digital counterparts are rather weak digital codes, suggests that even the best linear analog codes may not be that good after all. In the next section, we analyze the performance of linear analog codes on AWGN channels, propose strategies to design codes that are both MDRE and MDS, and evaluate how good linear analog codes really are.

## IV. ML DECODING AND DISTORTION

We start by looking into optimal decoders and distortion metrics. Consider the noisy reception $\mathbf{r}$ at the decoder. The maximum-likelihood (ML) decoder for a general analog code produces $\tilde{\mathbf{u}}$, where $\tilde{\mathbf{u}} = \arg \max_{\mathbf{u}} P(\mathbf{r}|\mathbf{u})$.

For a linear analog code operating on an AWGN channel, the ML decoder transforms to an unconstrained convex optimization problem:

$$\tilde{\mathbf{u}} = \arg \min_{\mathbf{u}} ||\mathbf{r} - \mathbf{G}^H \mathbf{u}||^2, \qquad (15)$$

which can be solved analytically by expressing the objective function as a convex quadratic function

$$||\mathbf{r} - \mathbf{G}^H \mathbf{u}||^2 = \mathbf{u}^H \mathbf{G} \mathbf{G}^H \mathbf{u} - 2\mathbf{r}^H \mathbf{G}^H \mathbf{u} + \mathbf{r}^H \mathbf{r}. \qquad (16)$$

The ML decision is obtained as
$$\tilde{\mathbf{u}} = (\mathbf{G}\mathbf{G}^H)^{-1}\mathbf{G}\mathbf{r}. \qquad (17)$$

We use mean square error (MSE) to evaluate the performance of an analog code on channels with additive noise $\mathbf{w}$, where the MSE distortion is defined as

$$\Delta = \int \left( P(\mathbf{u}) \int ||\tilde{\mathbf{u}} - \mathbf{u}||^2 P(\mathbf{w}) d\mathbf{w} \right) d\mathbf{u}, \qquad (18)$$

where $\tilde{\mathbf{u}}$ is the decoder output for source $\mathbf{u}$. For linear analog codes, because of the geometric uniformity, instead of evaluating over all the possible source vectors $\mathbf{u}$, the all-zero source vector can serve as the representative. Hence the MSE distortion can be simplified to:

$$\Delta = \int ||\tilde{\mathbf{u}}_0||^2 P(\mathbf{w}) d\mathbf{w}, \qquad (19)$$

where $\tilde{\mathbf{u}}_0$ is the decoder estimate for the all-zero codeword.

*Theorem 9:* Consider an $(n,k)$ linear analog code with encoder power gain $\Gamma$ operating on an AWGN channel with noise $\mathbf{w}$, where $w_i \sim \mathcal{N}(0, \sigma^2)$. The mean square error distortion $\Delta$ after ML coding is lower bounded by

$$\Delta \geq \Delta_{min} = \frac{k\sigma^2}{\Gamma} \qquad (20)$$

The lower bound is achieved by $s_0^2 = s_1^2 = ... s_{k-1}^2 = \Gamma$, where $\{s_0, s_1, ...s_{k-1}\}$ are the set of singular values of $\mathbf{G}$.

Proof: Without loss of generality, assume that the all-zero codeword is transmitted. Substituting $\mathbf{r} = \mathbf{w}$ and (17) in (19):

$$\Delta = \int ||(\mathbf{G}\mathbf{G}^H)^{-1}\mathbf{G}\mathbf{w}||^2 P(\mathbf{w}) d\mathbf{w}$$

$$= \int ||(\mathbf{G}\mathbf{G}^H)^{-1}\mathbf{G}\mathbf{w}||^2 \prod_{i=0}^{n-1} \left( \frac{1}{\sqrt{2\pi\sigma^2}} e^{-\frac{w_i^2}{2\sigma^2}} \right) d\mathbf{w}$$

$$= \int \frac{1}{(2\pi\sigma^2)^{n/2}} ||(\mathbf{G}\mathbf{G}^H)^{-1}\mathbf{G}\mathbf{w}||^2 e^{-\frac{\sum_i w_i^2}{2\sigma^2}} d\mathbf{w}$$

$$= \int \left( \mathbf{w}^H \mathbf{B}^H \mathbf{B} \mathbf{w} \right) \frac{1}{(2\pi\sigma^2)^{n/2}} e^{-\frac{\sum_i w_i^2}{2\sigma^2}} d\mathbf{w} \qquad (21)$$

where $\mathbf{B} = (\mathbf{G}\mathbf{G}^{-H})^{-1}\mathbf{G}$. Since $\mathbf{B}^H \mathbf{B}$ can be decomposed into the product of $\mathbf{A}^H \mathbf{D} \mathbf{A}$, where $\mathbf{A}$ is a unitary matrix, and $\mathbf{D}$ is a diagonal matrix, we can simplifye (21) to

$$\Delta = \int \left( \mathbf{w}^H \mathbf{A}^H \mathbf{D} \mathbf{A} \mathbf{w} \right) \frac{1}{(2\pi\sigma^2)^{n/2}} e^{-\frac{\sum_i w_i^2}{2\sigma^2}} d\mathbf{w}$$

$$= \frac{trace(\mathbf{D})}{k} \int \left( \mathbf{w}^H \mathbf{w} \right) \frac{1}{(2\pi\sigma^2)^{n/2}} e^{-\frac{\sum_i w_i^2}{2\sigma^2}} d\mathbf{w}$$

$$= trace(\mathbf{D})\sigma^2 \qquad (22)$$

The equality in (22) holds, because $\int \left( \mathbf{w}^H \mathbf{w} \right) \frac{1}{(2\pi\sigma^2)^{n/2}} e^{-\frac{\sum_i w_i^2}{2\sigma^2}} d\mathbf{w} = \sum_{k=0}^{k-1} E[w_i^* w_i] = k\sigma^2$ (i.e. the sum of the variance of $w_i$'s). Note that $\mathbf{B} = (\mathbf{G}\mathbf{G}^H)^{-1}\mathbf{G}$ is the psuedo-inverse of $\mathbf{G}$. Let $\{s_0, ...s_{k-1}\}$ be the set of singular values of $\mathbf{G}$, we have

$$trace(\mathbf{D}) = trace(\mathbf{B}^H \mathbf{B}) = \sum_{i=0}^{k-1} \frac{1}{s_i} \qquad (23)$$

To minimize $\Delta$ is then to minimize $\sum_{i=0}^{k-1} \frac{1}{s_i^2}$, subject to $\sum_{i=0}^{k-1} s_i^2 = k\Gamma$, which leads to:

$$s_1^2 = s_1^2 = ... = s_{k-1}^2 = \Gamma. \qquad (24)$$

Hence we have

$$\Delta \geq \frac{k\sigma^2}{\Gamma} \qquad (25)$$

*Corollary 10:* An MDRE code achieves the minimum bound of the MSE distortion on AWGN channels, and is therefore distortion optimal.

## V. Code Design

We showed that unitary codes, of which DFT and DCT/DST codes are special cases, are MDRE, and hence promise decent performance on AWGN channels. Since DFT and DCT/DST codes are also MDS (in terms of Hamming distance), they will also perform well on erasure channels or pulse channels. Question then arises as how and how easy it is to design linear analog codes that are both MDRE and MDS (and hopefully also simple). Below we provide a geometric view for linear analog codes and propose useful design rules.

A linear transform can de decomposed to a set of basic linear transformations: rotation, scaling, shearing, and reflection. Consider an arbitrary generator matrix $\mathbf{G}$, which can be singular value decomposed to $\mathbf{G} = \mathbf{A}\mathbf{D}\mathbf{B}$, where $\mathbf{A}$ and $\mathbf{B}$ are two square unitary matrices and $\mathbf{D}$ is a diagonal matrix whose diagonal elements are eigenvalues of $\mathbf{G}$. This suggests that an arbitrary linear transform can be implemented in three steps: rotating via the rotation matrix $\mathbf{A}$, followed by scaling via the scale matrix $\mathbf{D}$, and followed by a second rotation via matrix $\mathbf{B}$. Take a non-zero source vector $\mathbf{u}$, and we evaluate how the rotate-scale-rotate process may affect its Hamming weight and Euclidean weight. (For linear codes, weight of a non-zero sequence translates to the distance between a pair of sequences.) The vector $\mathbf{u}$ initially spans a $k$-dimensional subspace, which, when placed in an $n$-dimensional space, is like an $n$-dimensional vectors having $(n-k)$ zeros in the last $n-k$ dimensions. Since Hamming weight corresponds to the number of non-zero elements in the vector, scaling will not affect Hamming weight, but rotation will. Hence, as far as Hamming weight is concerned, one can safely assume that the scaling matrix is an identity matrix (no scaling on any dimension). What this implies in code design is that, for a given $(n,k)$ analog code, be it MDS or not, it is always possible to find a unitary code that produces exactly the same

Hamming weight spectrum. That is, a rotation matrix suffices to achieve the upper bound of the minimum Hamming weight.

Now to put (squared) Euclidean weight in perspective, it is clear that rotation becomes irrelevant and scaling takes the determining role. From our previous analysis of maximum squared Euclidean distance ratio and the minimum MSE distortion, the best scaling should be one that is uniform across all the dimensions. This is why codes whose eigenvalues of the $\mathbf{G}^H\mathbf{G}$ are identical are MDRE and simultaneously achieve the best minimum distance ratio and the best MSE distortion.

To conclude, the goals of optimizing Hamming distance and optimizing squared Euclidean distance do not conflict with each other in the context of linear analog codes. A good design can unify both metrics in one. For example, a carefully-selected matrix, such as that for an analog unitary code, achieves both MDRE and MDS bounds at one shot.

A related issue concerns repetition codes, which, as we have shown, are MDRE, and hence exhibit the same, best MSE performance on AWGN channels as any other MDRE codes. This result is verified by our simulations in Fig. 1. The y-axis represent the MSE distortion in log-scale, i.e. $\log_2(\Delta)$, where $\Delta$ is defined in (18). The simulation curves clearly show that repetition codes and DCT codes perform exactly the same on AWGN channels (in terms of MSE distortion), both noticeably better than the other two randomly generated analog linear block codes of the same parameters.

This result raises concerns on how good linear analog codes really are, especially considering that digital repetition codes are rather poor codes amongst digital codes. An important finding we wish to report here is: while linear digital codes are sufficient in achieving channel capacity (as exemplified by turbo codes and LDPC codes), linear analog codes are not; and to really perform well, analog codes must go nonlinear. Specifically, we show that in Fig. 2 two recently reported nonlinear analog codes (Baker's map codes and CAT codes) [10] [9], both of which significantly outperform MDRE analog codes (the DCT codes).

## VI. CONCLUSION

We have studied the theory of linear analog codes. We introduced the metric of minimum squared Euclidean distance ratio and mean square error distortion, established their respective upper bound and lower bound, and identified codes that simultaneously achieve these bounds. We also examined existing linear analog codes, and provided guidelines on designing codes that are both MDRE and MDS. While linear analog codes are (relatively) simple and analytically tractable, they are unfortunately weak. Unlike digital codes where linear codes are sufficient to combat AWGN, linear analog codes are inadequate and hence nonlinear analog mapping must be exploited for serious gains.

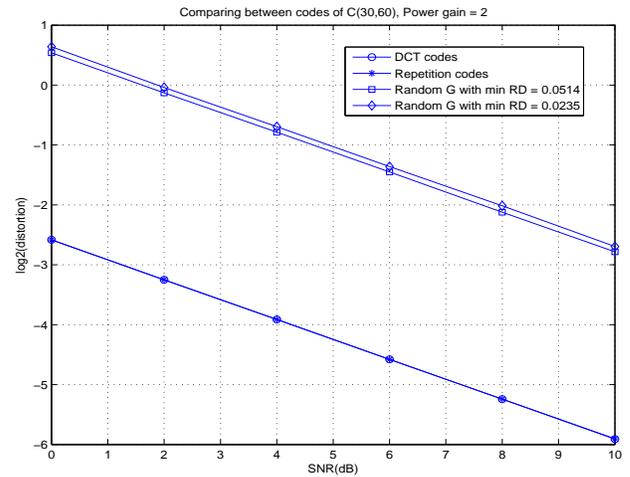

Fig. 1. Performance comparison of four $(60, 30)$ linear analog codes on AWGN channels: DCT, repetition, and two random analog codes with minimum distance ratio 0.0235 and 0.0514. All codes have encoding power gain 60. Source symbols are uniformly distributed over [-1,1]. SNR is measured in terms of $E_s s/N_o$, where $E_s$ is energy per channel symbol.

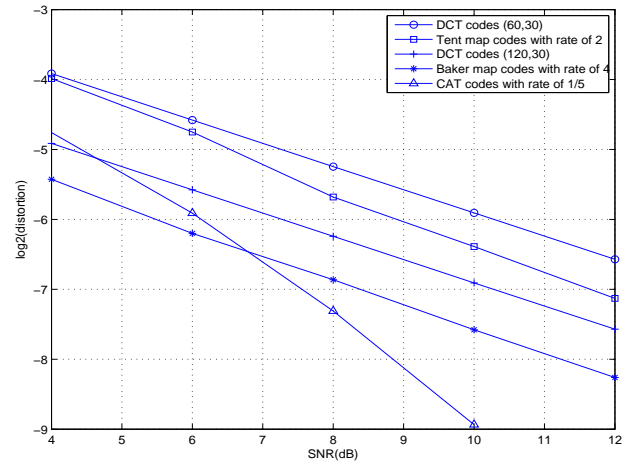

Fig. 2. Nonlinear analog codes can easily outperform linear analog codes.